\documentclass[%
 reprint,
 amsmath,amssymb,
 aps,
prl
]{revtex4-2}

\usepackage{graphicx}
\usepackage{dcolumn}
\usepackage{bm}

\usepackage[usenames,dvipsnames]{color} 	
\usepackage{floatpag} 					
\usepackage{footmisc} 					
\usepackage{booktabs}                   			
\usepackage{balance}                    			
\usepackage{tikz}

\makeatletter
\def\ps@pprintTitle{%
 \let\@oddhead\@empty 
 \let\@evenhead\@empty
 \def\@oddfoot{}%
 \let\@evenfoot\@oddfoot}
 \makeatother

\begin{document}


\title{Self-Arresting and Runaway Earthquakes:\\ Nucleation, Propagation, Gutenberg-Richter law and Dragon-King Events}

\author{Didier Sornette$^{1,2}$, Xueting Wei$^3$ and Xiaofei Chen$^{1,2}$}
\affiliation{%
$^1$  Institute of Risk Analysis, Prediction and Management (Risks-X), Academy for Advanced Interdisciplinary Sciences, Southern University of Science and Technology, Shenzhen, China\\
$^2$ Department of Earth and Space Sciences,  Southern University of Science and Technology, Shenzhen, China\\
$3$ School of Earth and Space Science and Technology, Wuhan University, Wuhan, 430072, China
}

\begin{abstract}
We develop a dissipation-based framework for earthquake rupture on homogeneous faults that explicitly separates the onset of unstable slip from the conditions required for self-sustained rupture propagation. This distinction explains the coexistence of self-arresting earthquakes and run-away ruptures (subshear and supershear events) observed in numerical simulations and empirical studies. We identify two distinct characteristic fault sizes: a nucleation radius controlling the instability of slip, and in general a larger propagation radius controlling whether an unstable rupture can be energetically sustained. Ruptures initiated above the nucleation scale but below the propagation scale spontaneously arrest. We further derive the Gutenberg-Richter law for self-arresting earthquakes by linking rupture physics to the fractal geometry of faulting. Finally, we interpret run-away ruptures as extreme events generated by an amplifying mechanism, consistent with the dragon-king concept. These results provide a unified physical basis for earthquake initiation, arrest, and seismicity statistics.
\end{abstract}

\maketitle


Earthquakes provide a natural laboratory for studying complex, out-of-equilibrium phenomena, displaying scale-invariant size distributions, multiscale fault structures, and emergent collective behavior, while remaining among the most consequential and unpredictable natural hazards.
The processes that lead to the onset and halting of an earthquake, as well as its preparatory stages, are still not fully understood.
Earthquake nucleation is commonly used as a broad term to describe the early stages of rupture development preceding a macroscopic earthquake. 
The concept of earthquake nucleation has a long and rich history in seismology, yet it remains conceptually ambiguous. As illustrated by numerous laboratory, observational, and theoretical studies \cite{OhnakaKuwahara1990,Beroza1995,AbercrombieMori94,Ide2019,Ben-DavidFine11,Ruina83,RiceGu83,Dieterich92,Ohnaka1993,Latouretal2013,McLaskey2019}, the term ``earthquake nucleation'' is commonly used to describe a wide range of early-stage processes preceding an earthquake, from slow aseismic creep and damage accumulation to accelerating slip and the onset of dynamic rupture. In doing so, the literature often uses the same terminology to refer to distinct physical notions, such as the loss of slip stability, the emergence of a dynamically propagating rupture, or the minimum size required for sustained rupture growth. This conflation has led to the widespread use of a single ``nucleation length'' to characterize rupture onset, implicitly assuming that the onset of instability and the ability of rupture to propagate are governed by the same physical mechanism.

Recent numerical and empirical evidence for self-arresting earthquakes \cite{XuSAR2015,WeiSAR2021,WenChenXu18,Galisetal23,XuMenyuan2023} challenges conventional interpretations of earthquake nucleation. These events initiate dynamically yet arrest spontaneously in the absence of material or geometric barriers, demonstrating that the onset of instability alone is insufficient to ensure sustained rupture propagation.

Here we present simple physical arguments and a minimal theoretical framework that explicitly separates the conditions governing the initiation of unstable slip from those controlling self-sustained rupture growth. This approach synthesizes disparate nucleation concepts found in the literature and provides a physically transparent basis for understanding how earthquakes may initiate, propagate, or self-arrest.

A central element of this framework is the recognition that earthquake rupture involves at least two distinct critical length scales, each associated with a different physical mechanism and operating at a different stage of rupture evolution. One length scale governs the local stability of slip and determines whether a small slipping patch becomes dynamically unstable. The other governs the energetics of rupture growth and determines whether an existing rupture front can be sustained. Although both scales are often referred to as ``nucleation sizes'' in the literature, they correspond to fundamentally different processes, and conflating them obscures the physical origin of distinct rupture outcomes.
In the following, we reserve the term ``nucleation'' for the stability-controlled onset of unstable slip, characterized by the nucleation radius $R_{\mathrm{nuc}}$. We refer to the second length scale, which controls whether rupture propagation can be energetically sustained, as the self-sustained propagation radius $R_{\mathrm{prop}}$.

Related nucleation mechanisms have been proposed at the asperity scale, notably the collective-detachment scenario \cite{deGeus2019Collective} where macroscopic slip is triggered when avalanches of asperity failures exceed a Griffith-like critical radius. More recently, velocity-weakening interfaces have been shown to generically exhibit coexisting power-law avalanches and rare system-spanning events that appear as statistical outliers \cite{deGeusWyart2022Scaling}, consistent with a dragon-king-like regime. While these studies emphasize disorder-controlled avalanches and microscopic nucleation mechanisms, our theory operates at a coarse-grained fault scale and explicitly distinguishes two characteristic lengths: a nucleation size controlling self-arresting events and a larger propagation threshold governing runaway ruptures. In contrast to asperity-controlled nucleation, we derive the Gutenberg-Richter law from the statistics of critical sliding patches and show that the largest earthquakes correspond to a distinct, amplification-driven regime consistent with the dragon-king framework.
 
We first study rupture initiation as a stability problem and introduce the nucleation radius $R_{\mathrm{nuc}}$.
Consider a slipping patch of characteristic size $R$ that undergoes a small additional slip increment $d\delta>0$. Two competing feedbacks determine whether this perturbation is stabilized or amplified: (i) elastic unloading (stabilizing) and (ii) frictional weakening (destabilizing).

\noindent
 \emph{Elastic unloading (stabilizing):} Because the surrounding medium relaxes as the patch slips, the shear stress acting on the patch decreases with slip. For small perturbations this can be linearized as
\begin{equation}
d\tau_{\mathrm{el}} = -\,k(R)\,d\delta,
\end{equation}
where $k(R)>0$ is the elastic stiffness (stress change per unit slip).
For a circular slipping patch of radius $R$, the elastic stiffness (stress change per unit slip) scales as
\begin{equation}
k(R)
\sim
C_k\,\frac{\mu}{R},
\end{equation}
where $C_k$ is a geometry-dependent constant of order unity and $\mu$ denotes the shear modulus of the elastic medium (with units of $Pa$), so that $k(R)$ has units of stress per unit slip ($Pa\,m^{-1}$). Ref.~\cite{Dieterich92} compiles values of $C_k$ ranging from $7\pi/24=0.92$ 
for a circular crack, to $2/3$ for a plane strain with constant shear stress, while smaller values are found for anti-plane strain.
Values of $C_k$ can exceed unity, for instance $C_k = 7\pi/16 \simeq 1.37$ for a circular shear crack,
because the effective stiffness depends on how ``slip'' is defined (e.g., average slip over the patch versus center or peak slip), so different crack idealizations naturally yield different numerical prefactors \cite{KeilisBorok1959,Shearer2009}.

\noindent
\emph{Frictional weakening (destabilizing):} Following  \cite{XuSAR2015,WeiSAR2021,WenChenXu18,Galisetal23,XuMenyuan2023}, we adopt a slip-weakening friction law, in which the frictional resistance decreases with accumulated slip, as a minimal and widely used description of fault weakening during dynamic rupture \cite{Ida72,Andrews1976a,Andrews1976b,Rice1980}.
For a linear slip-weakening friction law, the shear traction decreases linearly from the peak value $\tau_p$ to the residual value $\tau_r$ over a characteristic slip distance $D_c$. This can be written as
\begin{equation}
\tau(\delta)
=
\tau_p
-
\frac{\tau_p-\tau_r}{D_c}\,\delta,
\qquad
0 \le \delta \le D_c.
\end{equation}
The rate at which friction weakens with slip is therefore given by the slope of this relation,
\begin{equation}
\frac{d\tau}{d\delta}
=
-\frac{\tau_p-\tau_r}{D_c}
=
-\frac{T_u}{D_c},~~~~T_u \equiv \tau_p-\tau_r~.
\end{equation}
$T_u$ is the peak-to-residual (or breakdown) stress drop. 
The relevant quantity for stability is the magnitude of this slope, which defines the critical frictional weakening stiffness,
\begin{equation}
k_c
=
\left|\frac{d\tau}{d\delta}\right|
=
\frac{T_u}{D_c}.
\end{equation}
Physically, $k_c$ measures how rapidly the fault loses strength per unit slip once sliding initiates: larger $T_u$ or smaller $D_c$ correspond to more abrupt weakening and hence a greater propensity for instability.

Slip remains quasi-static (stable) only if the elastic stress reduction can ``keep up'' with (or exceed) the rate at which frictional resistance drops, so that a small slip increment does not generate a net driving stress increase. A convenient way to express this is to compare the stress change required by friction to remain on the weakening branch with the stress change supplied by elasticity. The net change in the \emph{driving} stress (applied minus resistance) due to $d\delta$ is
\begin{equation}
d\tau_{\mathrm{el}} - d\tau_{\mathrm{fr}}
=
\bigl(-k(R)\,d\delta\bigr) - \bigl(-k_c\,d\delta\bigr)
=
\bigl(k_c-k(R)\bigr)\,d\delta.
\end{equation}
Therefore:
\begin{itemize}
\item If $k(R) > k_c$, then $d(\tau_{\mathrm{el}}-\tau) < 0$: the driving stress decreases with slip, the perturbation is self-limiting, and slip is \emph{stable}.
\item If $k(R) < k_c$, then $d(\tau_{\mathrm{el}}-\tau) > 0$: the driving stress increases with slip, the perturbation is amplified, and slip becomes \emph{unstable}.
\end{itemize}

Hence, unstable slip (dynamic rupture initiation) occurs when elastic unloading is weaker than frictional weakening,
\begin{equation}
k(R) < k_c.
\end{equation}
Physically, this criterion states that, once sliding begins, the fault loses strength faster than the surrounding elastic medium can unload it, producing positive feedback and accelerating slip.
This yields the \emph{nucleation radius}
\begin{equation}
R_{\mathrm{nuc}}
=
C_k\,\frac{\mu D_c}{T_u} ~.
\label{h3ynbgq}
\end{equation}
The radius $R_{\mathrm{nuc}}$ is the critical size for the onset of unstable slip and controls the onset of unstable rupture:
a slipping patch of characteristic size $R > R_{\mathrm{nuc}}$ becomes dynamically unstable, 
so that any small slip perturbation grows into accelerated slip.
 Note that $R_{\mathrm{nuc}}$ is independent of the effective stress drop $T_e=\tau_0 - \tau_r$,
where $\tau_0$ is the applied shear stress $\tau_0$.

Expression (\ref{h3ynbgq}) appears counterintuitive at first sight: at fixed residual friction $\tau_r$, 
increasing the peak strength $\tau_p$ (and thus $T_u$) reduces the nucleation radius. This seems to suggest that a ``stronger asperity'' is easier to nucleate, which contradicts the naive expectation that stronger faults should be more stable. The resolution of this paradox lies in recognizing that rupture nucleation is a \emph{stability} problem governed by \emph{slopes}, not by absolute stress levels.
Indeed, a ``strong'' asperity in terms of peak stress can be dynamically fragile. High peak strength delays the onset of slip, but high weakening rate accelerates the onset of instability once slip occurs. Nucleation is controlled by the latter, not the former.
A larger peak-to-residual stress drop does not make rupture harder to nucleate; instead, it increases the abruptness of frictional weakening once slip starts. This enhanced brittleness overwhelms elastic stabilization at smaller scales, leading to a smaller nucleation radius.

\paragraph{Conditions for self-sustained propagation and introduction of the propagation radius $R_{\mathrm{prop}}$.}

After instability is triggered at the nucleation scale $R_{\mathrm{nuc}}$, rupture may either arrest or continue to expand, depending on whether elastic energy release is sufficient to sustain front propagation.

Consider a circular slipping region of radius $R$ that has grown beyond the nucleation radius and is therefore undergoing dynamic slip. The surrounding elastic medium is subjected to an effective stress drop
\begin{equation}
T_e = \tau_0 - \tau_r,
\end{equation}
where $\tau_0$ is the applied shear stress and $\tau_r$ is the residual frictional stress. The elastic energy decrease associated with the existence of this slipping region scales as
\begin{equation}
E_{\mathrm{el}}(R)
=
-\frac{1}{2}\,\frac{T_e^2}{\mu}\,\eta\,\frac{4\pi}{3}R^3,
\end{equation}
where $\mu$ is the shear modulus and $\eta$ is a geometric factor of order unity,
identical to the inverse of previous constant ($\eta=1/C_k=1.09$) for a circular crack \cite{Eshelby1957,Eshelby1959}.
The incremental elastic energy released when the rupture radius grows from $R$ to $R+dR$ is
\begin{equation}
-\frac{dE_{\mathrm{el}}}{dR}
=
\frac{1}{2}\,\frac{T_e^2}{\mu}\,\eta\,4\pi R^2.
\end{equation}

Since the newly slipped area is $dA = 2\pi R\,dR$, the elastic energy release rate per unit area is
\begin{equation}
G(R)
=
\eta\,\frac{T_e^2}{\mu}\,R.
\label{fdnejht}
\end{equation}
This quantity governs the energetics of \emph{front advance} once a rupture exists.

As the rupture expands beyond the nucleation scale and dynamic slip develops, further growth requires dissipation of frictional energy at the rupture front. This dissipation is governed by the fault's weakening behavior and is concentrated within a finite cohesive (process) zone, which must be supplied with sufficient energy for the rupture to advance into previously intact material.
As before, we assume a linear slip-weakening friction law with peak-to-residual stress drop $T_u$ over a critical slip distance $D_c$. The fracture (or frictional dissipation) energy per unit area associated with full weakening thus reads
\begin{equation}
G_c
=
\int_0^{D_c} \left[\tau(\delta)-\tau_r\right]\,d\delta
=
\frac{1}{2}T_u D_c.
\label{ju3hga}
\end{equation}
This quantity $G_c$ epresents the localized energy dissipation required within the cohesive (process) zone to weaken the fault from peak to residual strength and thereby allow the rupture front to advance into fully weakened, sliding material. It constitutes the energetic ``entry cost'' for converting intact fault area into a slipping fault.
  
A rupture front can propagate in a self-sustained manner only if the elastic energy release rate exceeds the required frictional dissipation energy:
\begin{equation}
G(R) \ge G_c~.
\label{trhjtun3thgq}
\end{equation}
Within the general framework of out-of-equilibrium thermodynamics, rupture propagation can be viewed as a form of pattern formation sustained by a continuous flux of energy through the system. The rupture front constitutes an organized, spatially localized structure that can persist and advance only if the mechanical energy supplied by the surrounding elastic medium is sufficient to compensate for irreversible dissipation. In this sense, a propagating rupture is a dissipative structure in the sense of nonequilibrium thermodynamics.
The elastic energy release rate $G(R)$ represents the rate at which free (recoverable) mechanical energy is made available by extending the rupture front. Frictional weakening and inelastic processes within the process zone irreversibly convert this energy into heat and other microscopic degrees of freedom, producing entropy. The associated dissipation is quantified by the frictional dissipation energy $G_c$, which measures the minimum irreversible energy cost required to create new slipped area. Self-sustained propagation is therefore possible only when the supplied energy flux exceeds the dissipation rate required to maintain the advancing front (\ref{trhjtun3thgq}).
From a thermodynamic perspective, this condition expresses the requirement that entropy production localized at the rupture front can be continuously supported by the energy flux from the elastic reservoir. If $G(R) < G_c$, the energy input is insufficient to maintain the dissipative structure, entropy production cannot be sustained at the front, and rupture propagation arrests. Conversely, when $G(R) > G_c$, the excess energy may be radiated as seismic waves or stored transiently as kinetic energy, while the minimum dissipation $G_c$ ensures irreversibility of rupture advance.
In this view, rupture initiation corresponds to a loss of stability of an equilibrium or quasi-static state, whereas rupture propagation corresponds to the maintenance of a nonequilibrium steady state at the rupture front. The distinction between nucleation and propagation thus mirrors the broader distinction in nonequilibrium thermodynamics between instability thresholds and sustainment conditions for dissipative structures \cite{NicolisPrigogine1977,CrossHohenberg1993,Haken1983}.

Substituting the expressions (\ref{fdnejht}) and (\ref{ju3hga}) in (\ref{trhjtun3thgq}) defines the minimal radius for self-sustained propagation 
\begin{equation}
R_{\mathrm{prop}}
=
\frac{\mu\,T_u\,D_c}{2\eta\,T_e^2}.
\label{jumjn}
\end{equation}
This radius does \emph{not} control rupture initiation
which is determined by $R_{\mathrm{nuc}}$ (\ref{h3ynbgq}).
Self-arresting ruptures on homogeneous faults thus emerge naturally from the competition between elastic energy release and irreversible frictional dissipation in a driven system. The critical radius $R_{\mathrm{prop}}$ marks the boundary between arrested ruptures ($R<R_{\mathrm{prop}}$) and self-sustained rupture growth ($R>R_{\mathrm{prop}}$).

Note that the residual friction work $\int \tau_r\,d\delta$ is not localized at the rupture front and does not scale purely with the newly created area $dA$. It is dissipated progressively in the rupture wake, behind the front, as slip accumulates. Because it is not an incremental cost required to advance the rupture front into new material, it does not enter into the  frictional dissipation  energy per unit area associated with full weakening 
that enters into the propagation condition (\ref{trhjtun3thgq}).

Even under sustained far-field driving, the local stress available at the rupture front may decrease as the slipped region grows, due to elastic unloading, stress shadowing, or radiative losses. As a consequence, the local energy release rate $G(R)$ can fall below the dissipation threshold $G_c$ for sufficiently small or intermediate rupture sizes. In this case, rupture propagation becomes energetically unfavorable and the event self-arrests, without requiring any geometric or material barriers.

\paragraph{Synthesis: initiation, propagation, and self-arrest.}

The analysis developed above reveals that earthquake rupture evolution is governed by two distinct characteristic length scales, each rooted in a different physical mechanism. The nucleation radius $R_{\mathrm{nuc}}$ marks the loss of stability of quasi-static slip and controls whether a localized slipping patch becomes dynamically unstable. In contrast, the propagation radius $R_{\mathrm{prop}}$ is an energetic threshold that controls whether an already unstable rupture can continue to grow in a self-sustained manner. These two radii are therefore not alternative definitions of the same concept, but complementary scales operating at successive stages of rupture evolution.

The rupture outcome depends on the relative size of the initial slipping or rupturing patch $R_0$ with respect to these two scales. If $R_0 < R_{\mathrm{nuc}}$, elastic unloading dominates frictional weakening and slip remains stable or aseismic. If $R_{\mathrm{nuc}} < R_0 < R_{\mathrm{prop}}$, instability is triggered and dynamic rupture initiates, but the elastic energy release is insufficient to overcome the fracture energy required for continued front advance, leading to a self-arresting rupture.  Increasing the critical slip distance $D_c$ tends to promote arrest, whereas increasing the available stress drop $T_e$ suppresses it. 
Only when $R_0 > R_{\mathrm{prop}}$ does rupture growth become energetically self-sustained, allowing runaway propagation.

Self-arresting ruptures therefore occupy a genuine intermediate regime that emerges naturally when stability-controlled initiation is distinguished from energetically controlled propagation. This regime exists only when the energetic threshold exceeds the nucleation threshold, $R_{\mathrm{nuc}} < R_{\mathrm{prop}}$.
Substituting expressions (\ref{h3ynbgq}) and (\ref{jumjn}) yields the condition
\begin{equation}
{T_e \over T_u} = {\tau_0 - \tau_r \over \tau_p-\tau_r} < \frac{1}{\sqrt{2\eta C_k}} \simeq 1~.
\label{hwbhb2gq}
\end{equation}
For cracks, $\eta C_k=1$, so inequality~(\ref{hwbhb2gq}) is always satisfied, 
since the peak stress $\tau_p$ is usually larger than the applied shear stress $\tau_0$.
This is consistent with the common situation in which the nucleation scale 
$R_{\mathrm{nuc}}$  is smaller than the propagation scale $R_{\mathrm{prop}}$. More generally, across a wide range of loading conditions, fault geometries, and degrees of slip heterogeneity, this ordering is expected to hold in many cases, although it need not be universal, as shown below. 
 
\begin{figure}
	\centering
	\includegraphics[width=9cm]{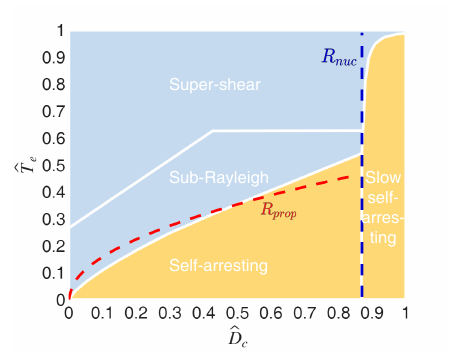}
	\caption{Phase diagram in the plane $[ {\hat D}_c ;  {\hat T}_e]$, defined by expressions (\ref{eq:ar1}),
	 taken from  \cite{WeiSAR2021} on which the curve ${\hat T}_e = 0.5 \sqrt{{\hat D}_c}$ is 
	superimposed (dashed red line) as obtained from the derivation leading to equation (\ref{eq:pwyyf8f89}) with $\eta=2$. 
	The vertical blue dashed line corresponds to $\hat{D}_c = 1/C_k$ (equation (\ref{eq:pwyyf8f89}))
	with $C_k \approx 1.15$, which separates the slow and fast self-arresting earthquakes. 
}		
	\label{trhtgbvqf}
	\end{figure}
	
Recent numerical studies by Xu et al.~\cite{XuSAR2015} and Wei et al.~\cite{WeiSAR2021} provide a particularly relevant context for testing the predictions of the present energy-dissipation theory. Using fully dynamic simulations based on the boundary integral equation method, these authors systematically investigated rupture evolution on homogeneous planar faults governed by linear slip-weakening friction. Their focus was on how rupture behavior depends on both the properties of the slip-weakening law and the size of the initially perturbed (nucleation) asperity.

In these studies, rupture was artificially nucleated within a finite circular patch of radius $R$, and its subsequent evolution was tracked without imposing barriers or heterogeneities. By varying the breakdown stress drop $T_u$, the effective dynamic stress drop $T_e$, the critical slip distance $D_c$, and the nucleation size $R$, Xu et al.~\cite{XuSAR2015} and Wei et al.~\cite{WeiSAR2021} identified four distinct rupture regimes: (i) subshear (or sub-Rayleigh for strike-slip rupture) propagation, (ii) supershear propagation,  (iii) self-arresting ruptures that initiate dynamically but stop spontaneously and (iv) slow self-arresting ruptures that stop within the nucleation zone.
Importantly, self-arrest occurs despite homogeneous fault properties and in the absence of any external arresting mechanism, highlighting that rupture initiation and rupture sustainment are controlled by different physical conditions.

To organize their results, Xu et al.~\cite{XuSAR2015} and Wei et al.~\cite{WeiSAR2021} introduced a phase diagram expressed in terms of two dimensionless parameters,
\begin{equation}    \label{eq:ar1}
        \hat{T}_e := \frac{T_e}{T_u},
        \qquad
        \hat{D}_c := \frac{D_c}{R}\,\frac{\mu}{T_u},
\end{equation}
where $T_u$ is the breakdown stress drop, $T_e$ is the effective (dynamic) stress drop relative to the residual stress, $D_c$ is the critical slip distance (typically in the range $0$--$3~\mathrm{mm}$ in their simulations), $R$ is the effective radius of the nucleation asperity, and $\mu \approx 30~\mathrm{GPa}$ is the shear modulus of the surrounding elastic medium. When expressed in this nondimensional form, the numerical results collapse onto well-defined regime boundaries separating  fast and slow self-arresting ruptures from sustained subshear and supershear events.

This numerical phase diagram provides a natural benchmark for the two predictions (\ref{h3ynbgq}) for the nucleation radius $R_{\mathrm{nuc}}$
and (\ref{jumjn})  for the propagation radius $R_{\mathrm{prop}}$ of the energy-dissipation framework.
Rewriting expressions (\ref{h3ynbgq}) and  (\ref{jumjn}) in terms of the nondimensional variables $\hat{T}_e$ and $\hat{D}_c$ (\ref{eq:ar1}) gives
\begin{equation}    \label{eq:pwyyf8f89}
\hat{D}_c = 1/C_k~ (\ref{h3ynbgq}) ~~~~,~~~~~
\hat{T}_e
=
\sqrt{\frac{\hat{D}_c}{2\eta}}~(\ref{jumjn}) ~.
\end{equation}
The first equation $\hat{D}_c = 1/C_k$ corresponds to the vertical blue dashed line in Fig.~\ref{trhtgbvqf} for $C_k \approx 1.15$,
which separates the slow and fast self-arresting earthquakes. 
The independence of the dimensionless critical slip distance $\hat{D}_c$ from $\hat{T}_e$
provides a nontrivial validation of the theory.
Moreover, the relation $\hat{T}_e = \sqrt{\frac{\hat{D}_c}{2\eta}}$
predicts a sharp boundary between self-arresting and sustained ruptures. 
This theoretical boundary, shown as the red dashed line in Fig.~\ref{trhtgbvqf} for $\eta=2$
can be directly compared with the numerical phase diagrams of Xu et al.~\cite{XuSAR2015} and Wei et al.~\cite{WeiSAR2021}. 
The close agreement between theory and simulations provides strong support for the present framework.

Remarkably, the theory captures not only the transition threshold but also the overall shapes of both phase boundaries (blue and red dashed lines). This agreement is obtained with only two adjustable parameters, $\eta \approx 2 \pm 0.5$ and $C_k \approx 1.15 \pm 0.1$. Substituting these values into inequality~(\ref{hwbhb2gq}) yields $ \frac{1}{\sqrt{2\eta C_k}} \approx 0.7 \pm 0.1$,  implying that $R_{\mathrm{nuc}} < R_{\mathrm{prop}}$ only for 
$ \hat{T}_e ={T_e \over T_u} = {\tau_0 - \tau_r \over \tau_p-\tau_r} \lesssim 0.7\pm 0.1 $.
 Consequently, for $\hat{T}_e \gtrsim 0.7\pm 0.1$, one has $R_{\mathrm{nuc}} > R_{\mathrm{prop}}$,
 which corresponds to the upper region of slow self-arresting rupture in the phase diagram. In this regime, any rupture with 
 $R>R_{\mathrm{nuc}}$ inevitably evolves into a runaway rupture, either sub-Rayleigh or supershear.
While the precise numerical value of the boundary is sensitive to the simplified geometry underlying the theoretical argument, the overall consistency reinforces the robustness of the proposed phase-diagram framework.
 
\paragraph{Derivation of the Gutenberg-Richter distribution for self-arresting earthquakes.}

Having shown that the dissipation-based framework provides a quantitative explanation for the transition between self-arresting and sustained ruptures, we now explore its implications for earthquake statistics. In particular, we derive the Gutenberg-Richter distribution for self-arresting earthquakes with rupture radii smaller than the self-sustained propagation scale $R_{\mathrm{prop}}$ given by expression~(\ref{jumjn}). Our approach links the fractal geometry of faulting to rupture nucleation physics through the dependence of the critical size on the dynamic stress drop.

In the Earth's crust, the propagation scale $R_{\mathrm{prop}}$ is not expected to be uniform but instead to vary spatially, reflecting heterogeneity in material and frictional properties. Variations in the elastic modulus $\mu$, breakdown stress drop $T_u$, effective dynamic stress drop $T_e$, and critical slip distance $D_c$ naturally lead to a distribution of local propagation thresholds. As we show below, the combination of this heterogeneity with the scale-invariant structure of faulting yields the observed Gutenberg-Richter law for self-arresting earthquakes.
Given that $R_{\mathrm{prop}}$ is inversely proportional to the square of $T_e$,
the largest contribution to the variations of $R_{\mathrm{prop}}$ at large values comes from 
the variations of the dynamic stress drop $T_e$ at small values.
Let us assume that $T_e$ is distributed according to some spatial distribution $p(T_e)$, which is regular for small $T_e$.
The simplest forms are either a constant or an exponential distribution for its probability density function (pdf) for small $T_e$'s
and give the same asymptotic result and the same Gutenberg-Richter distribution. Let us take
$p(T_e) = {1 \over T_0} e^{-{T_e \over T_0}}$,
which is normalised to $1$ for $T_e$ spanning from $0$ to infinity (this upper boundary is of no consequence for the distribution of 
$R^c$ at intermediate and large values).
Expression (\ref{jumjn}) together with the exponential form of $p(T_e)$ leads to the pdf of $R_{\mathrm{prop}}$ given by
\begin{equation}    \label{eq:awtr2et}
p(R_{\mathrm{prop}})  = \sqrt{{1 \over 2\eta} {T_u \over T_0}  {\mu \over T_0} D_c}~
      e^{- {1 \over T_0}\sqrt{{T_u \mu D_c  \over 2\eta  R_{\mathrm{prop}}}}}~{1 \over (R_{\mathrm{prop}})^{3 \over 2}} ~.
      \end{equation}
This implies $p(R_{\mathrm{prop}}) \propto  {1 \over (R_{\mathrm{prop}})^{3 \over 2} }$ for  $R_{\mathrm{prop}} \geq {T_u  \mu D_c \over 2 \eta T_0^2}$.
Consider a fault network of span $L$. The number of independent patches of size $R$ on this fault network is proportional to $\left({L \over R}\right)^{D_f}$, 
where $D_f$ is the fractal dimension of the fault network.
Thus the total number of nucleation domains that are in the self-arresting rupture regime is the sum over all possible 
critical radii $R^c$ weighted by its pdf:
\begin{equation}    \label{eqwgr:awtr2}
      p(R) =   \left({L \over R}\right)^{D_f}  \int_{R}^{+\infty} dR_{\mathrm{prop}} ~p(R_{\mathrm{prop}})   \simeq  {1 \over R^{D_f+{1 \over 2}}}~.
\end{equation}
The lower bound in the above integral expresses the fact that a given asperity of size $R$ is in the self-arresting rupture regime
if it is smaller than its local critical radius.  The pdf of the disk radius $R$ translates into the pdf of seismic moment $M_0$
using the standard scaling relation $M_0 \propto R^3$. Using the conservation of probabilities under a change of variable
$p(R) dR = p(M_0) dM_0$ yields
\begin{equation}    \label{eqwgr2ta15tg1tr2}
      p(M_0)   \simeq  {1 \over M_0^{1+{D_f -{1 \over 2} \over 3}}}~.
\end{equation}
Kagan and Knopoff \cite{KaganKno80} report a value $D_f=2.2$, which should be taken as a lower bound since it is obtained from 
the two-point correlation function of the distribution of earthquakes. Davy et al. \cite{DavySor2_90} give a value $D_f=1+1.7=2.7$ from 
analog laboratory experiments of the India-Asia collision. Sornette \cite{SornetteGeilo91} reviews a series of measurements by various authors that give 
$D_f$ in the range $2.5-2.7$. Taking $D_f=2.5$ recovers exactly the $2/3$ exponent of the Gutenberg-Richter distribution 
of seismic moments, equivalent to the $b$-value equal to $1$
when expressed in terms of magnitudes.
This derivation informs that the Gutenberg-Richter distribution holds for self-arresting earthquakes.
In the present framework, the Gutenberg-Richter distribution derives from the combination of the fractal nature of faulting and the physics of earthquake nucleation
via the dependence of the critical nucleation size as a function of the dynamical stress drop.
Variations of the $b$-value could be associated with different fault network properties with distinct fractal dimensions.

\paragraph{Interpretation in terms of the theory of dragon-kings.}

Within the framework developed above, earthquakes naturally separate into two qualitatively distinct classes depending on whether rupture growth is self-limited by dissipation or amplified by positive feedback. Self-arresting ruptures occur when instability is triggered but elastic energy release remains insufficient to sustain propagation. Their growth is therefore intrinsically limited by local energetics and dissipation, leading to a population of events whose sizes are controlled by local material properties and heterogeneity. In contrast, once the propagation threshold is exceeded, rupture growth becomes self-amplifying: elastic energy release increases with rupture size, dissipation can be continuously sustained, and rupture expansion is no longer limited by local frictional weakening but only by the encounter of strong geometric or structural barriers.

We propose that these non-self-arresting, run-away ruptures correspond to instantiations of the concept of dragon-kings \cite{DK09,DK12}, i.e., extreme events that are both exceptionally large (``kings'') and generated by mechanisms fundamentally different from those governing the bulk of the population (``dragons''). In this interpretation, the subshear and supershear ruptures that propagate until arrested by major geometric barriers constitute the dragon-kings, whereas the subcritical self-arresting ruptures represent the ``normal'' events distributed according to the Gutenberg-Richter law~(\ref{eqwgr2ta15tg1tr2}).

This classification follows directly from the nucleation and propagation physics developed here. Self-arresting earthquakes are numerous because they require only the onset of instability, which can occur over a broad range of conditions, but they remain limited in size by dissipation. In contrast, run-away ruptures require the additional and more restrictive condition that elastic energy release exceeds dissipation at all scales, making them intrinsically rarer. Once nucleated, however, their sizes are no longer set by local frictional parameters but by the spatial organization of fault networks and the statistics of large-scale barriers. As predicted by dragon-king theory \cite{DK09,DK12}, these extreme events should therefore follow a magnitude distribution distinct from that of the smaller, self-arresting earthquakes.

This mechanism-based interpretation of run-away ruptures as dragon-kings naturally complements recent empirical and statistical studies that test the dragon-king hypothesis in earthquake catalogs. These works identify some large earthquakes as statistically significant outliers to the Gutenberg-Richter distribution and argue that they arise from distinct physical rupture mechanisms rather than from the continuation of scale-invariant seismicity \cite{DK-Li25,DK-Li-HaichTang}.

This work is partially supported by the National Natural Science Foundation
of China (Grant no. U2039202, T2350710802), Shenzhen Science and Technology
Innovation Commission (Grant no. GJHZ20210705141805017), and the Center for
Computational Science and Engineering at Southern University of Science and
Technology.

\balance
\bibliographystyle{apsrev4-2}
\bibliography{bibliography}

\end{document}